\documentclass[]{article}

\usepackage{amsmath}
\usepackage{graphicx}
\usepackage{amsfonts}
\usepackage{ae}
\usepackage{graphics,epsfig,latexsym}
\input epsf

\usepackage{lineno}
\usepackage{setspace}

\usepackage{epstopdf}

 \usepackage[hyperindex=true,colorlinks=true,pagebackref=false,linkcolor =black,citecolor=blue,plainpages=false,pdfpagelabels,breaklinks=true]{hyperref}
\usepackage{natbib}
 
\usepackage{t1enc,verbatim}
\usepackage{times}

\begin{document}


\linenumbers

\title{\textbf{Living \emph{is} information processing: from molecules to global systems.}}
\author{Keith D. Farnsworth$^{1}$, John Nelson$^{1}$ \& Carlos Gershenson$^{2}$ \\
\small $^{1}$ School of Biological Sciences, Queen's University Belfast, UK, BT9 7BL \\
 \small \{k.farnsworth,John.Nelson\}@qub.ac.uk\\
\small $^{2}$ Instituto de Investigaciones en Matem\'{a}ticas Aplicadas y en Sistemas, \\
 \small Universidad Nacional Aut\'{o}noma de M\'{e}xico.\\
 \small A.P. 20-726, 01000, M\'{e}xico, D.F., M\'{e}xico\\
 \small cgg@unam.mx
}
\date{}

\maketitle

\doublespacing

\begin{abstract}
We extend the concept that life is an informational phenomenon, at  every level of organisation, from molecules to the global ecological system. According to this thesis:  (a) living is information processing, in which memory is maintained by both molecular states and ecological states as well as the more obvious nucleic acid coding; (b) this information processing has one overall function - to perpetuate itself; and (c) the processing method is filtration (cognition) of, and synthesis of, information at lower levels to appear at higher levels in complex systems (emergence).  We show how information patterns, are united by the creation of mutual context, generating persistent consequences, to result in `functional information'. This constructive process forms arbitrarily large complexes of information, the combined effects of which include the functions of life. Molecules and simple organisms have already been measured in terms of functional information content; we show how quantification may be extended to each level of organisation up to the ecological. In terms of a computer analogy, life is both the data and the program and its biochemical structure is the way the information is embodied. This idea supports the seamless integration of life at all scales with the physical universe. The innovation reported here is essentially to integrate these ideas, basing information on the `general definition' of information, rather than simply the statistics of information, thereby explaining how functional information operates throughout life.

\end{abstract}


\newpage

\section{Introduction: what is life?}

The question `what is life' is one of the oldest in philosophy, deeply mysterious and still fascinating. Not only is it fundamental to biology, it has challenged and extended physics, metaphysics, the human sciences of medicine and psychology, the arts and even spiritual thinking. But efforts to answer the question have generally been constrained by disciplinary boundaries or within an organizational scale of life, leading to several apparently separate answers. The aim of this paper is to unite these by considering life as a whole, simultaneously at every organizational level (from molecule to global ecosystem). This integration uses the concept of life as information processing for a unifying principle.

During the second half of the 20th century, the paradigm that `life is chemistry' \citep{KORNBERG:1991rz} was especially influential in understanding living processes at the sub-cellular level. As increasingly complicated networks of molecular interactions were recognised, the need for a formal understanding of their organizational structures developed into systems biology, which now extends beyond the cell \citep{Kohl:2010fm}. At the same time, but largely unrelated, theoretical ecology developed into a form of cybernetics: the study of self-regulating systems, moving chemical substances through networks of populations and communities. The complex networks of the cell's biochemistry were paralleled by complex webs of interactions among organisms: the elaborate complexities of the `-omics' were matched by those of biodiversity as we realised that the estimated 15 million species (8.7 million eukaryotic \citep{Mora:2011eu} plus 6 million prokaryotic \citep{Curtis:2002uk}) are all connected to one-another in networks of community interactions. Observing that these complex networks may be two manifestations of a common feature of life, we now propose a unifying model in which interactions among molecules, cells, organisms and populations all amount to information processing through a hierarchy of functional networks - molecules in cells, cells in organisms and organisms in communities, which compose the biosphere. This model, which extends recent developments in systems biology \citep{Maus:2011fk} is intended to integrate through all life over its entire history. 

Biologists know that information is crucial to life, pointing to its role in DNA for maintaining the design of organisms over repeated generations and an understanding of information in protein structure has a long history \citep[see e.g.][]{Yockey1958}. A cybernetic view goes further to claim that information processing, carried out in the medium of biological chemistry, is what life actually \emph{is}. By information processing we mean any logical combination of information having the result of producing information and we shorten this to `computation'. The idea that `living is computing', pioneered by theorists such as \citet{Galtin1972} has been popularised by \cite{BRAY:1995ph,Bray2009}, but so far, it has been contained within cellular biochemistry (with computation by neural networks the obvious exception). Our aim is to show how well the whole of life can be viewed in this way as an integrated information processing system: all cells working together. This view seamlessly connects with the concept of information as one of three elemental components of existence (with space/time and matter/energy) which has grown within physics over the past several decades, accompanied by a new philosophical position which places information at the core of determining reality (termed `Informational Structural Realism' by \citet{Floridi2003}). Every aspect of life may be regarded as a product and elaboration of the physical world, clearly made of the same matter and energy, ordered in space and time as is every physical system. What makes life special is not the material brought together to take part in living, it is the functional information that orders matter into physical structures and directs intricate processes into self-maintaining and reproducing complexes. In the information model of life, this definitive process (termed autopoiesis by Maturana and Varela \citeyearpar{Maturana1980} consists of a system of structural elements continually replacing themselves to maintain the living system by following a program of instructions that both makes their information-rich structure and is instantiated within it. Significantly, this fundamental feature of life is true at every organizational scale, not only at the cellular level.

\section{Information Concepts}

According to the `diaphoric definition of data' \citep{Floridi2003,Floridi:2005kx}, a binary bit (the unit of information) is a single difference. For example, a digital monochrome image of $k$-pixels instantiates no more than $k-1$ differences. When the image carries a meaningful picture, it instantiates fewer than the maximum number of differences, so can be compressed by recording only the differences where black changes to white. The maximally compressed image instantiates $k-n$ bits ($n \geq 1$) and this is termed the Algorithmic Information Content (AIC) \citep{Chaitin1990}. The same applies not just to representations, such as images, but to real physical objects: a compressible pattern of differences makes an object what it is. This refers not to a description, but to the physical object itself, giving a definition of physical information as a pattern of difference: the algorithmic information embodied by an object so as to give it form. Information in this sense, selects the elementary particles of the object and specifies the locations of these in space and time (under quantum-theoretic constraints). The minimum description of the object is the AIC embodied in both this physical configuration of particles and the nature of each (Pauli's exclusion principle ensures these are different). On a technical note, AIC is known not to be strictly computable \citep{Li2008}, but an effective substitute is available in the Computable Information Content for empirical studies needing to compute it \citep[see e.g.][]{Menconi2005}.

For most practical purposes, in describing an object, we would consider higher levels of abstraction, such as a pattern of atoms, molecules, cells, tissues, or components, etc.. Again, for most practical purposes, we are concerned not with the total AIC instantiated in an object, but with the \emph{functional information content} (FIC), which is the part of AIC which can cause a persistent change of information in any part of the system. As an illustration, two seemingly identical metal keys will be different in detail (at the small scale), but may both function to open the same lock: their functional information defines their shape as fitting the lock. This is obviously pertinent to biology through the lock and key analogy of messenger molecules, but also describes functional equivalence among all kinds of biological molecules; among cells of the same type and state in the body; and among organisms of the same function in an ecosystem. FIC can be quantified, as demonstrated at the nucleotide level by \cite{Jiang:2010lr}, who calculated it as the minimal amount of genomic information needed to construct a particular organism. We hope to apply this idea to structures of biological information, other than the genetic. 

In the field of Biosemiotics, pieces of functional information are regarded as symbols \citep[see][]{Favareau2009}, but we wish to focus on the functioning of information, rather than its communication. For this, we take the idea of \emph{function} from \cite{Szostak:2003kl}, seeing it as what makes systems, including biological ones, operate, in the sense of an operational explanation of function \citep{Neander2011}. The definition of  `function' has been debated among philosophers for several decades and deserves some attention here. \cite{Cummins75} proposed that function is an objective account of the contribution of a system component to the `capacity' of the system. Crucially, for Cummins, the capacity (meaning capability) of a system is explained in terms of the capacities of the components it contains, and how they are organised. This concept explicitly matches the understanding that functional information is to be found in the component parts and the way they are organised into a whole. But it has been criticised, especially for its permitting what appear to be unintended consequences as functions (a frequently cited example being that dirt in a pipe may `function' as a valve \citep{Griffiths93}). One of the solutions to this, at least for organisms, is to recognise that natural selection tends to eliminate potential functions of components if they do not contribute to the biological fitness of the system of which they are a part. This qualification was taken up by \cite{Neander91}, by developing a biologically-based  etiological theory. Whilst appealing, this cannot be used for all biological systems, such as ecological communities, for which evolution by natural selection has not been established, so to be general, we are forced back to the systemic theories of function. However, Darwin's theory is a special case of a more general principle of selection in which the attribute of persistence is the superset of biological fitness \citep[e.g.][]{Kauffman93}. Thus we tentatively offer a definition of function that is systemic and in the spirit of established etiological definitions, but not reliant on Darwin's theory. It is that any attribute A of a component C of a system S that causes an effect E such that S persists longer or in a wider range of conditions than without it, is a functional attribute of C. Then the functional information instantiated by C is that which establishes A, leading to the persistence in form of S, hence the persistence of information instantiated by S. 

Szostak's \citeyearpar{Szostak:2003kl} mathematically amenable definition allows for a quantification of the effectiveness with which information enables a system to perform non-random actions; at least one of which will be self-replication. From here on, we shall use the general term `\emph{effective information}' for that which causes a persistent change, so has an effect in the wider system and reserve the term `\emph{functional information}' for effective information which plays a role in supporting life. We note that at the specifically nucleotide level, since evolution selects for function, non-functional information will be lost from biological systems over evolutionary time (this was demonstrated by Schneider's \citeyearpar{Schneider.ev2000} `evolutionary program'). However, non-functional information is continually introduced by random processes, especially at higher (e.g. ecological) levels, so non-functional `noise'  may be expected and should be discounted in the quantification of FIC. 


We take as axiomatic that information is instantiated in matter through the particular arrangement of its components in space and time. This arrangement defines a unique relationship among the components, which can only instantiate information if it is stable and therefore persists as a configuration in space over a line in time. When two or more such configurations are brought into association, there is a combined arrangement, which if persistent, also instantiates information: that of both components \emph{plus}  that of their association. The Shannon information \citep{Shannon1948} of the combined configuration is given by the product of probabilities of each component configuration (less any mutual information). Thus the `surprise' in finding this new whole is in general greater than that for each of its component parts. Nested construction of increasingly complicated configurations of matter may proceed this way and thereby constitute an increase in information content in the Shannon sense \citep{Shannon1948}. Most significantly, when configurations combine into stable forms, they do so by presenting context for one another: the information of each is functional information for the other, enabling greater function than that of the sum of parts. 

The functional meaning of information was defined conceptually by \cite{MacKay1969} who referred to information as ``a distinction that makes a difference'' and later \citet{Bateson1972} more famously called information ``a difference that makes a difference'', this idea was then taken up by \cite{Hopfield1994}. In this interpretation, information is defined through its interaction with something (including other information) to create a non-random effect, hence it is context dependent. \citet{Bates2005}, quoting earlier works, defines information as: ``the pattern of organization of matter and energy". This definition peculiarly addresses effective information. Patterns of organization are the alternative to randomness: patterns show either order (characterised by symmetry) or complexity (broken  symmetry).  \citet{Schrodinger1944} realised that symmetrical order was insufficient to account for the genetic information coding life, concluding that it must be in some aperiodic (non-symmetrical) molecule (well before the discovery of DNA). The required organized aperiodicity is commonly known as `complexity'; a defining characteristic of which is a high capacity for effective information.  \citet{Adami:2000oq} subsequently showed how all biological systems are complex systems in this scientific sense.

These concepts are brought together in Figure 1 which shows three levels of information concept in the formation of life. On level 1, physical information is understood as the result of an improbable (following Shannon's insight) and persistent configuration of energy and/or matter in space and time. In level 2, effective information is defined through consequence: a contextual relation is made among at least two such configurations (now considered as information and termed `infons'). This synthesis through mutual context is exemplified by a lock and key enzyme interaction. Level 3 takes this further to capture the idea that a large number of contextual interactions structure an assembly of infons into a complex system; exemplified by a molecular network inside a cell. Not shown is the hierarchical concept that such systems can be the component parts of super-systems, enabling an unbounded construction of nested complexity, in which information at higher levels, but not present at lower levels, can be defined and measured as emergent \citep{GershensonFernandez:2012}. That is the way life appears under observation, exemplified by the notional hierarchy in figure 2 and table 1. 

\begin{figure}
  \includegraphics[width=1.0\textwidth]{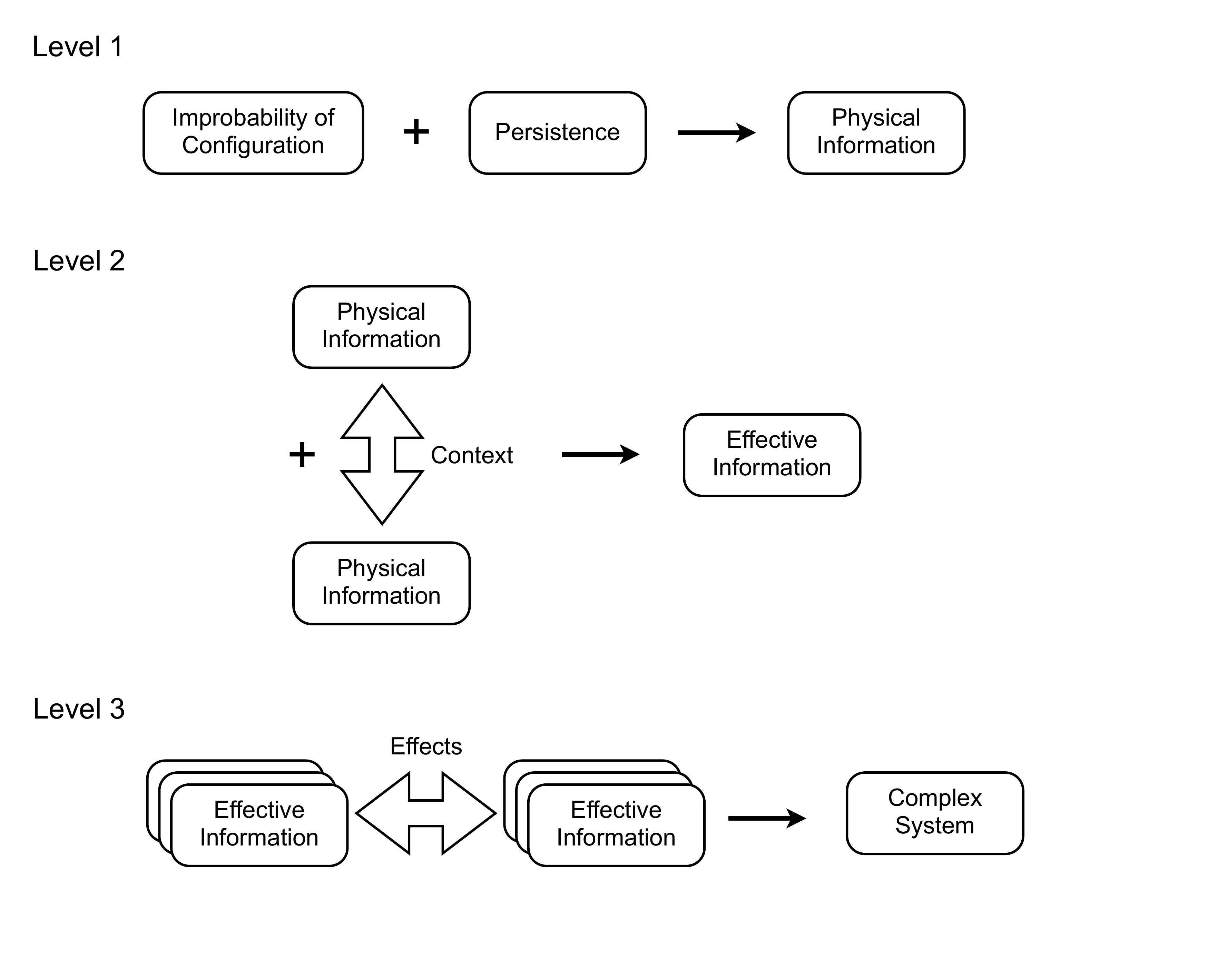}
\caption{Three levels of information concept explained in the text: at level 1, information is a pattern of difference; at level 2, information becomes effective through context and at level 3, `packages' of effective information combine, affecting one another to form a complex system that computes.}
\label{fig:1}       
\end{figure}

\newpage

\subsection{Order from disorder: self-assembling structures}

According to statistical mechanics, the organization of a system is the result of filtering, i.e. selecting a particular configuration of system component states from all possible configurations and this filtering is equivalent to investing the system with information, in the Shannon sense \citep{Shannon1948} of reducing the probability of its configuration. When the resulting organization causes sustainable self-assembly, using active filtration from the wider environment, the system may be said to live. 

It is most parsimonious to assume that the components of matter needed to constitute living organisms were originally distributed in perfect randomness (disorder). Apparently, life alone creates life, but before it appeared for the first time, individually persistent (non-transitory) stages of ordering among collections of molecular components must have occurred. It is broadly understood that this develops through the spontaneous emergence of `order out of chaos'  \citep{Foerster60, Prigogine84, Kauffman93}---in which chaos then referred to disordered randomness. This natural evolutionary phenomenon, which obeys the second law of thermodynamics, is very general. It amounts to the selection of more stable configurations from a set of random configurations, simply by virtue of their stability conferring greater persistence. Darwin's evolution by natural selection is a particular instance of this process, which also applies to resonance phenomena and crystal formation. 
 
Life orders matter, but differs from a crystal in the following critical respects: (a) life is a dynamic pattern not a static one; (b) it is not regular, but rather is complex, meaning that it cannot be summarized in a short piece of information and (c) it manipulates its environment so as to make its persistence more likely. The vortex (e.g. a whirlpool) is an often cited example of a non-living system which displays some of these properties. It maintains itself as a dynamic pattern of matter, even though its constituent parts are  constantly changing: molecules which pass through in a moment are replaced by others, but the pattern and therefore the structure-forming information is maintained. This is an example of a `dissipative structure' defined and recognized as self-organizing by Prigogine \citeyearpar{Prigogine77}. By continually exchanging matter and energy with their environment, these dynamic structures are able to continually `dissipate' entropy, with the effect of concentrating information. This information is instantiated in the form of the structure. Crucially the essence of these dissipative systems is organizational information, not substance, and the information they maintain has the special property of being that which is necessary for the self-maintenance. 

Given the required material components and thermodynamic conditions, we see that information in the form of a pattern in matter can emerge spontaneously and maintain itself as long as these conditions allow. The next step is to ask if it can also create the components and maintain the conditions it needs to do this in a changing environment. If any pattern can achieve that feat, then it will be able to reproduce and ensure its persistence far longer than thermodynamics would otherwise allow. The ability of a system (any arrangement of matter) to remake itself is termed autopoiesis and this has been identified as one of the two necessary capabilities of anything living \citep{Maturana1980}. The other is cognition, more precisely, the detection and selection of particular elements from an environment of many random elements, which is a kind of information processing. Bitbol and Luisi \citeyearpar{Bitbol:2004tv} showed that autopoiesis and cognition are separately necessary conditions for life, not inseparably linked as apparently first thought by Maturana and Varela \citeyearpar{Maturana1980}. They illustrated their point with reference to the autopoietic fatty acid cells, which Zepik et al. \citeyearpar{Zepik:2001ta} showed to achieve reproduction and self-maintenance by homeostatic processes autonomously generated from within.  From this work, it became clear that for a system to live, it must have at least the following three properties: autopoiesis, cognition and an unbroken boundary to define its limits \citep{Bitbol:2004tv}; this latter stops the ingredients of life from diffusing apart, rendering life's chemical reactions too rare to work as a whole. In practice, all known living systems are cellular\footnote{Though some biologists may include viruses.} and indeed, the cell tegument has never been broken since the beginning of life---it has only been divided by repeated fission. Division among organisms is just an elaboration of division among cells. In this sense all life from its beginning, is unified as a set of cells, related through replication; all creating order from disorder, by cognition and autopoiesis. 

The result of this long history of accumulating functional information in a population of diverging cell lines is illustrated in Figure 2 where the major developments are illustrated. By specializing into specific types, cells have found ways to more effectively live: colonies of specialist cells forming into the distinct tissues of separate organisms, organized into ecological communities, interacting, to the point of regulating the earth's geochemistry through a homeostatic network. All of this amounts to information processing---selecting molecules from the environment, ordering matter and controlling flows of matter and energy. The information needed to perform these functions is found distributed among the molecules within every cell: not just in nucleotides, but in all the proteins and messenger molecules, their interactions and locations in space. However, seeing life as a whole in space and time, from the first single cell to all extant life, implies an integrated system, for which hierarchical levels represent merely \emph{observed} abstractions of organisational structure \citep[see][]{salthe1985}. Considering the whole living system from notional levels of biochemistry at the bottom to global ecosystem at the top, we may regard all but one of the levels in table 1 to be a model, the single exception being organisation into cells. Hierarchy theory recognises constraints imposed by higher levels on the lower, but also the constraint of possibilities from lower levels upwards. We understand the need for bounded cells as one of those possibility constraints and therefore see cells as the one exception - they are not merely a model level but one in the reality of life's organisation.

\subsection{Biological systems as effective information}

It is evident that the minimum functional information needed to constitute life is large (the smallest non-virus functional information content calculated so far is  $2.86.10^6$ bits for \textit{Holarctica} \citep{Jiang:2010lr}). By current consensus, life emerged as an entropy-dissipating pattern which created and maintained a boundary through which trapped molecules were able to selectively interact with the wider environment \citep{Morowitz1992,SmithMorowitz2004}. This cognitively filtering system also reproduced itself by growth and fission and all extant life followed via evolution \citep{RobertsonJoyce2010}. The resulting proto-cell was a complex dynamic system in which  information was held, not just in the component molecules, but also in the interactions among them. These interactions instantiated functional information because the molecules gave context to each other, thereby filtering out specifically functional interactions from the whole range of possibilities. 

The cytoplasmic contents of cells are spatially structured so that the time and place of interaction is a necessary determinant of their effect. Because molecular components are distributed in a specific spatial pattern, their collective behavior is  extended to form regions of coordinated, but different action over space. This instantiates functional information in spatial relations so that simple unitary systems (e.g. enzyme interactions) combine to exhibit complex behaviors which \emph{appear} to be the product of more complicated components. The apparently spontaneous emergence of new information \citep{GershensonFernandez:2012}, is in fact the revelation of that spatio-temporal information already present in the distribution of components and the network of signaling paths among them (a phenomenon first described by \citealp{Turing1952}). Any spatio-temporal information (coding the positions of system elements in time and space) that contributes to the emergent behaviors of the whole system, is effective information, and in life this is maintained by autopoiesis. When a more complicated system is created from simple units in this way, it results in a new unit, the combination of these being the next tier in an hierarchy of complexity. It is by this nested hierarchical construction that the enormously complex machinery of life is brought into being.  

Information is therefore not just stored in nucleotides: it is the whole biological system that embodies effective information, hence biocomplexity as a whole is the storage of effective information in living nature. \cite{Valentine2003} realised this and emphasised that biological complexity exists as a set of hierarchical levels, as we illustrate in table 1 (adapted from \cite{FarnsworthLyashevska2012}). Spontaneous creation of effective information from complex order is a signature property of such hierarchies: every level spontaneously \emph{emerges} from the one below \citep{Adami:2000oq, Lorenz2011129} - all the way up to global ecosystems. For this reason, even a complete description of genetic information fails to account for the full complement of effective information in life, which is why seed-banks and zoos are no substitute for community conservation, as noted intuitively by \citet{Lee2004} and \cite{Cowling2004}. Indeed, `living information' is only fully instantiated in dynamic, active systems capable of flexibly responding to environmental conditions. A common example is the gene-regulatory network, which apparently extracts maximum autopoietic complexity by functioning near criticality \citep{Balleza:2008ix}, where information content is maximised  \citep{GershensonFernandez:2012}.

\subsection{Quantifying Functional Information}

\citet{FarnsworthLyashevska2012} classified the total information content of any system into two distinct components: $I_{tot} = I_F + I_R$, where $I_F$ is the functional information and $I_R$ is the random information. Each of these terms can be quantified by the Algorithmic Information Content \citep{Chaitin1990} if the term can be isolated. $I_F$ could, in principle, be quantified by the `Effective Complexity' \citep{GellMann1996,GellMann2003} , defined as the minimum description length of regularities, but only given prior knowledge about the regularities \citep[see][for an expansion of this criticism]{McAllister2003}. To describe life as information, we need a way to identify $I_F$ without such prior knowledge, recognising that effect only results from the interaction of information and its context. 
In the special case of genomes, this is relatively trivial since almost all the information present is functional \citep{Schneider.ev2000}. For quantification, \cite{Jiang:2010lr} defined `effective information' as that part of the genome which is the minimum needed to reconstruct the organism. This meant estimating the functional (coding) fraction of the genome and (manually) compressing it to form the equivalent Algorithmic Information Content. In an application of Boltzmann's entropy concept at the genetic level, \cite{Szostak:2003kl} defined 'functional information', in terms of a gene string, as $-\log_2$ of the probability that a random sequence will ``encode a molecule with  greater than any given degree of function'' - in other words a design brief, without implying a designer. In the case of genes, this `function' may be thought of as the biochemical activity (for example a digestive enzyme's catalytic rate) of whatever molecule is produced from reading the nucleotide sequence. This design-brief concept was developed to the ecosystem level of organisation by \citet{FarnsworthLyashevska2012}, who interpreted it as a set of ecological functions and related functioning to the information content of food-web networks.

\section{The natural history of information processing} 

We have argued that life is a dynamic process of filtering and communicating information. The processing of information (computation) occurs in all cases of changing, combining and directing information. Thus computation is a natural, continuous and ubiquitous process (see \cite{Denning2007}). However, it is important to distinguish between (a) universal computing, which can represent any computation in symbols that may be 'programmed' and (b) fixed computing in which the hardware and software are interdependent, so that only a narrow range of computational tasks may be performed (this point is discussed by \citet{Hopfield1994}). Life is very much in the latter category (though since the brain is one of its products, this is not universally the case). Complex system computation is now a well established model in behavioral ecology, describing many aspects of social organization (reviewed by \citealt{Camazinebook2001}). Other kinds of computation performed by life include information replication, ordering and re-ordering of form and cybernetic system control, each of which will be briefly illustrated below. In each case, computation occurs on a distributed network \citep{Gershenson:2010b}, rather than through the linear Von-Neumann architecture of the familiar digital computer. Whether looking at molecular networks or ecological communities, we see that natural computation is composed of cybernetic feedback loops arranged functionally so that the system gains in persistence. That these loops exist is not a surprise, since any random interconnection of quantities may contain loops and many physical processes do. As control circuits they may generate positive feedback, often leading to quick destruction, or negative feedback leading to stability, and hence more likely to persist in their changing environment. In fact, since control of this kind enhances persistence, natural selection favours cybernetic systems (with negative feedback) above others and we may find this kind of computation practically inevitable. However, a network solely composed of negative feedback fixes on a particular equilibrium, so may be insufficiently flexible to perform the processes of life \citep{Kauffman93}. Since a mix of positive and negative feedback loops can create a dynamic and adaptable system of `state-cycles' in the narrow `critical' regime between catastrophe and order (exemplified by random Boolean networks with high link densities) this has been proposed as an essential feature of living systems by \cite{Kauffman93} and we now look for evidence of these in significant developments of biological organisation (Figure 2).


\begin{figure}
  \includegraphics[width=1.0\textwidth]{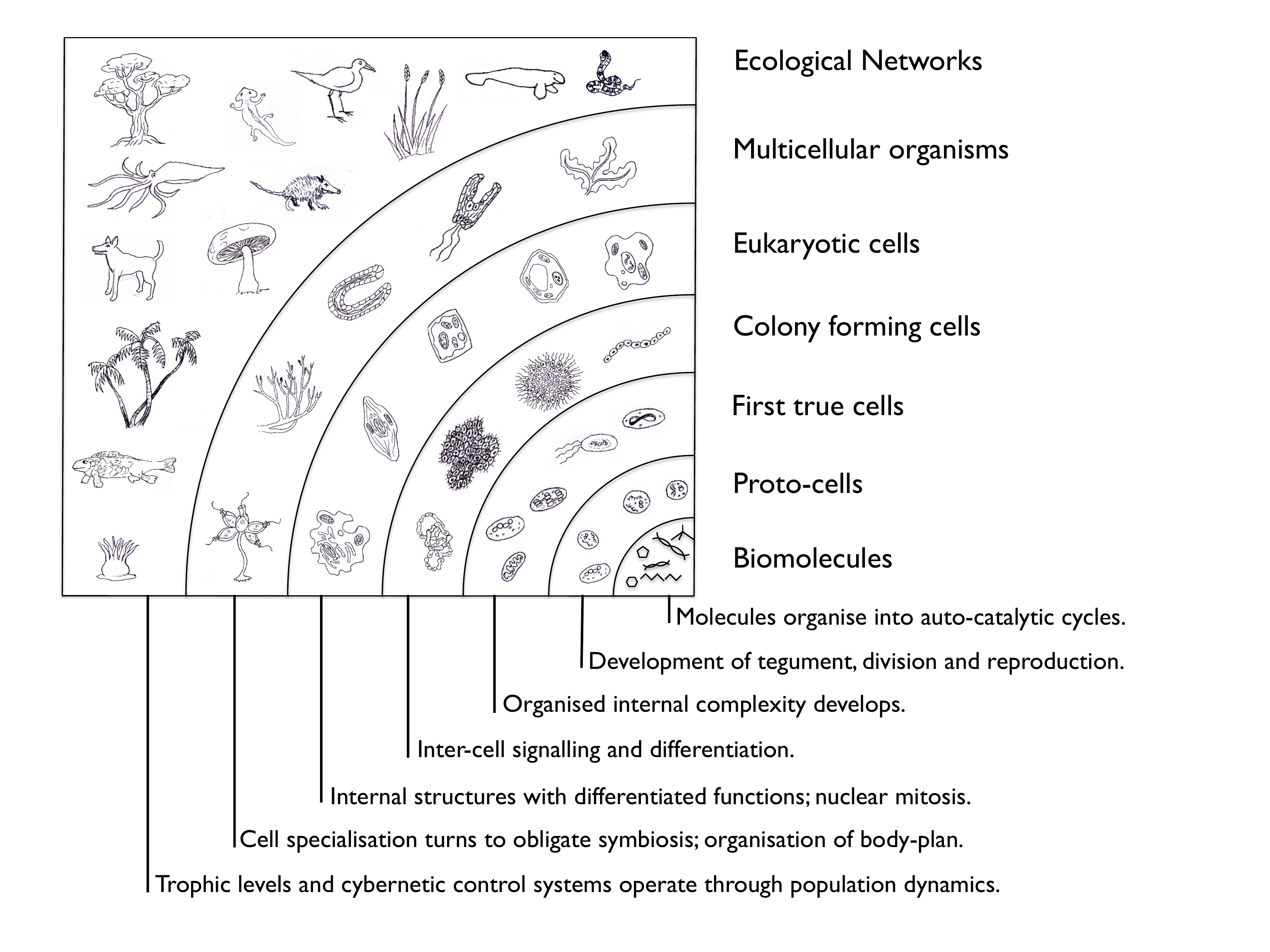}
\caption{Hierarchical self-assembly of complex systems: the increase in computational complexity through the history of life on earth, often associated with a major transition e.g. from prokaryotic to eukaryotic life-forms, or the development of cell-signalling networks or ecological networks. Note that concentric rings indicate expansion of complexity, rather than a chronological sequence: all inner layers exist concurrently at each level.}
\label{fig:2}       
\end{figure}

\subsection{Computing through cell-signaling networks}

If living is the self-sustaining \emph{coordination} of chemical reactions, does this  suggest a coordinating manager? The nucleus was once thought to be the `command centre' of the eukaryotic cell, but observations of cells behaving normally for months after enucleation show that the information processing needed for most activities is cytoplasmic \citep{Goldman1973}. It would be better to think of the nucleus as the `hard disk' of the cell, since here (for the human) the `blueprints' for at least  47 thousand different proteins  \citep{Orchard:2005qd} are stored and transcribed, together with editable instructions about when to make them. The `algorithms of living' are run on these proteins which act in ways analogous to transistors and other electronic components, in complex networks, as described by \cite{Paton1998}.

The model of cellular information processing as analogue computation (e.g. Rodbell \citeyearpar{RodbellM1995}) was inspired by the cybernetic theory of Norbert Wiener \citeyearpar{Wiener1948}. In this model, external chemical messages  (first messengers) are first `discriminated' (by the receptor) then `transduced' (by a G protein) and finally amplified (by an effector enzyme) to produce an intracellular signal (the second messenger)---a sequence that can be summarized as perception. This second signal typically initiates a complex sequence of interconnected changes which may alter the internal chemistry of the cell, change the response to other first messengers, and even selectively alter gene expression \citep{Cairns1988}. Such cascades of molecular response form dynamic networks that carry and process information {\citep{Lehn:1990}, analogous to artificial neural networks. Chemical switches are implemented by the allostery of proteins, especially enzymes, acting as `transistors' in the network circuitry \citep{BRAY:1995ph}. Furthermore, activated proteins do not simply diffuse to collide with their targets. Cytoplasm is a well organized and densely crowded environment in which the reaction cascades are localized by `scaffold' proteins, reminiscent of the electronic circuit board. For example, the protein kinase enzyme, type II PKA may be fixed to either the plasma membrane, the cytoskeleton, secretory granules, or the nuclear membrane by anchoring proteins \citep{Scott1992}. The effect is not only to position this signaling protein close to its intended target but also to determine the local molecular environment (context) which may profoundly influence the effect. Such protein networks are built and repaired following the DNA blueprint, which as we have just noted, may itself be altered by the cytoplasmic computation. Thus, proteins dynamically send, receive and respond to informational signals in complex and dynamically changing networks of both negative and positive feedback, which, collectively interacting with stored DNA-information, form the behavior of the cell and this is readily interpreted as molecular computation. 

\subsection{Replicating information}
Biological reproduction is an information transfer (communication) phenomenon, from parent(s) as the transmitter to daughter(s) as the receiver. This biological communication requires a high standard of accuracy, since the information being transmitted is very nearly all functional  \citep{Schneider.ev2000}. Given this view of reproduction as efficient semantic communication, it was a surprise to realise that the length of the nuclear genome bears no relation to organism complexity \citep{Gregory2001,Valentine2003}.  Since the complexity of a system can be defined as the minimum amount of information needed to describe (or reproduce) it, one possible reason is that species differ in the amount of error-mitigating repetition their genomes carry. As well as this, the DNA of almost all organisms harbours a zoo of information parasites (selfish DNA - \citet{Orgel1980}) and their remnants, making up a large part of what was historically referred to as `junk DNA' when its function was unknown. Transposable elements form the majority of this repetitive information \citep{Wessler:2006ls}. It is now thought that many of these `transposons' originated as endogenised retro-viruses \citep{BowenJordan2002}: parasites that have been co-opted into functional symbiosis under regulation by the host \citep{Veitia:2009zz}. This legacy of non-host information accounts for a large part of the huge variation in genome size among eukaryotes, where multiple copies of information parasites are found. However, the relationship between nuclear genome size and organism complexity is still an open question. 

Given our understanding of emergence and the formation of functional information from mutual context, we can see that not all of the functional information is to be found in nuclear DNA. So whilst physically, it is the genes that are replicated in biological reproduction, context-dependent relationships among them constitutes functional information that is carried along with the replication. Gene regulatory networks (GRNs) \citep{DavidsonLevin2005} are the most significant information complexes to extend beyond nuclear DNA and are composed of context-dependent relationships among infons, rich in both negative and positive feedback. Again, these networks are readily modeled as computational systems \citep{Kravchenko-Balasha:2012tb} and their role in determining body-plan through epigenetic phenomena points to a possible correlation between GRN complexity (hence information content) and organism complexity. 

\subsection{The eukaryotic revolution}
Following pioneering work by \citet{Margulis1970}, endosymbiosis is the front-running theory explaining the origin of eukaryotic cells and this well illustrates the increase of function brought about through the creation of mutual context among infons (level 2 in Figure 1). The advantage of eukarotic cells over prokaryotic is the specialisation of metabolic, anabolic and reproductive machinery. The component parts  collectively become more efficient by (a) individually concentrating on a smaller task and (b) sharing the products. The fundamental reason this narrowing of tasks improves effectiveness is that it reduces the information requirement for performing all necessary tasks. If we think of a cell as a machine performing $n$ processes; it needs storage capacity enough to instantiate the algorithms for all $n$ tasks. prokaryotic cells have rather limited storage capacity (determined by their AIC), so cannot afford a very sophisticated algorithm for every task they have to perform - they are limited in effectiveness by their information capacity limit. When a cell incorporates others, it increases its storage capacity and permits a distribution of tasks among specialist components, each of which can devote the whole of their limited storage capacity to carrying a sophisticated and efficient algorithm for a single task. It is also necessary to include the communications and sharing among the specialist components, so some algorithm space is devoted to this. The exchange among individual components forms a network of control computation, which on a larger scale constitutes a complex system (level 3 in Figure 1). 

\subsection{Cell types and body-plan complexity}

Information's role in ordering of form is most apparent in the building of multi-cellular organisms. Cells come in a large variety of forms, with hierarchical morphotype structure and developmental lineages \citep{Valentine2003}. The number of distinct cell types in a single organism is taken as an indicator of its complexity \citep{Carroll:2001fe} and varies among metazoan phyla from 3 (Myxozoa) to 210 (human) having steadily increased through evolutionary time \citep{VALENTINE:1994kn}. This indicates a gradual accumulation of biological complexity, and therefore functional information,  as life-forms have radiated and cell specialisation has apparently increased. Despite that, \citet{Hinegardner83a} concluded that ``evolution since the Cambrian appears to have involved few major increases in biological complexity", as Valentine \citeyearpar{VALENTINE:1994hi} argued, the basic body-plans of all extant phyla were established by the end of the Cambrian explosion (520 My ago). The apparent contradiction may be explained by proliferating patterns of gene expression, rather than the creation of new genes; this being one of the central hypotheses of evolutionary development biology \citep[see][]{Valentine2003bk}. Such proliferation of patterns and the consequent radiation of organism-forms is the result of ordering and re-ordering of functional information. Different cell-types are created by regulating the expression of different genes in the total genome---simpler organisms suppress the expression of a higher proportion of their developmental genes than do complex ones \citep{Davidson:2001aq}. Thus, the morphological complexity of an organism is determined by the regulatory machinery which selects genetic expression during the development of an organism. The number of cell types is one computed `output' of gene regulatory networks and gives a very rough indication of functional information content. A trend in modeling body-plan  regulatory networks, represents them in a way analogous to artificial neural networks \citep{Geard2005a}, clearly interpreting morphogenesis as computation. This suggests a means of quantifying the functional information of body plans by experimentally (\emph{in silico}) examining variants of formative gene-networks and recording the resulting morphometric diversity.  

\subsection{Cybernetic computation by ecological communities}

Darwin's metaphor of a `tangled bank' suggests a bewildering complex of interactions among whole organisms \citep{Montoya2006}, but natural computation is rarely, if ever, explicit in ecological models. Information processing in ecological communities is less clear than in cells and organisms because ecosystems usually lack obvious boundaries and their functions are usually considered, not at the system level, but at the population level, where cybernetic control is not apparent. However, some recent developments pave the way for this to change; both in describing the information content of communities and in understanding them as self-regulating complex systems. 

The study of biodiversity provides a starting point to finding the functional information content at the ecological level. Using the idea that difference is the basis of information \citep{Floridi:2005kx}, diversity (which by definition counts total difference) becomes a measure of information content. Traditionally, biodiversity describes the number of different species and perhaps the evenness of their abundances in an ecological assembly, using metrics inspired by Shannon's information theory \citep[see][]{magurran2004}. More recently, broader definitions recognize diversity at every level in the biological hierarchy (table 1), and ecologists may now refer to genetic and functional diversity as equally necessary for specifying biodiversity \citep{Lyashevska2012}. Ecological communities can be regarded as the vaults of information capital, in the form of molecular structures; networks and pathways; cell types; tissues and organs, whole organisms and community interactions \citep{FarnsworthLyashevska2012}. However, ecologists still refer to organizational scale through informal terms: for example `alpha' and `beta' diversity, which are arbitrarily defined phenomena of classification \citep{Tuomisto12010} over probability distributions \citep{McGill2011,Nekola1999}. 
Whilst hierarchical nesting of complex systems is explicitly recognised by multi-level modelling in sub-cellular biology, the strength of formal description this provides has yet to enter ecology \citep[see][]{Faeder:2011lr}.

Descriptive approaches can be developed into conceptual models by changing the focus towards the network through which organisms interact \citep[as in][]{NORTON:1992tx}. The basic components for such models are  available in the special case of predator-prey interactions \citep[e.g.][]{Dunne:2002qq} (though models of other material and informational -- e.g. genetic -- flows are less well developed). Given a network description such as a food-web, \cite{FarnsworthLyashevska2012} showed how the functional information approach may be applied at the ecological level. They  systematically dismantled a network model of the Northeast Atlantic fish community, at each stage measuring its productivity, to find a relationship between complexity and function, which provided a measure of the marginal change in function with network (algorithmic) information content. Food-webs are but a partial description of ecosystems, which necessarily include chemical, energy and information flows. Being relatively simpler, microbial networks are more amenable to this fuller description. The recent development of functional and genetic network models in microbial ecology \citep[e.g.][]{Zhou:2010fj} gives us a stepping stone between sub-cellular networks and community level computation. Significantly, microbial colonies preceded the close association of eukaryotic cells to form multi-cellular organisms \cite{Lepot2008}, yet specialisations among microbe species imply the same need for self-regulating interactions as is found in organismal physiology. By definition, an isolated microbial community must be autopoietic and as specialisation among constituent species develops, so must flows of coordinating information work to compute the community, via complex-system emergence.  

\subsection{Information processing as an integrated whole.}

The computation performed by ecological networks is both broader and narrower than that of a Turing machine (a system following a sequence of logical operations defined by \citealp{Turing1936}). It is broader because inputs are processed continuously, the outputs are produced continuously and because processing is sensitive to the environment (in a Turing machine, processing is blind to all but the initial inputs until a halt condition is reached, releasing the output). It is narrower because the computation is equivalent to running a particular model: a model of the system under control, following the injunction of \citet{ConantAshby1970}, that ``Every good regulator of a system must be a model of that system''. Interactions between the biotic and abiotic spheres of the ecosystem are regulated by controls on chemical flows, driven by the processing of materials by life \emph{in aggregate}: the sum of individual selection and processing actions amounts to a regulation of the whole ecosystem. Including the flows of nutrients such as nitrogen and phosphorus in ecological network analysis \citep[e.g.][]{Ulanowicz1999} takes us a step closer to the biochemical analogy of within-cell computation \citep{Ulanowicz1980}. Since molecules continually flow through ecosystems, just as they do in the cell, we can identify the process of constant renewal of ecosystem structure (the network) as autopoiesis, this time referring to all life in aggregate. The phenomenon of constant renewal by recycling material, driven by transforming high to low entropy energy, accumulated over all life on earth, is the foundation of the Gaia hypothesis \citep{Lovelock1974}. The total of global ecological processes may be interpreted as a network computer, whose input is the physical and chemical environment of the planet and the output is a computed adjustment of these to maintain equilibrium. Seen this way, life is a computer running a model of itself in order to control its interior state so as to perpetuate itself in a changeable environment. This view, which goes beyond cybernetic self-regulation to reveal autopoietic computation, is closely allied to a growing thermodynamic understanding of living processes in which energy accountancy is integrated with informational interpretations  \citep[e.g. in][]{Smith:2008hy}. For example, the accumulation of hierarchical complexity, so characteristic of life, has been demonstrated to follow from thermodynamic efficiency \citep{Wicken1979, Annila2008, Annila:2009kl} as has the tendency for hierarchical complex structures to regulate their internal and external environments through information processing \citep{Kaila:2008pb, Karnani:2009rq}. 

\section{Implications}

The information perspective shows life to be (a) continuous with the abiotic universe and (b) the consequence of a spontaneous increase in complexity through repeated combination of formative patterns such that they give context and thence function to one-another. Chemistry is the result of this process at the atomic scale and life is a branch of chemistry that is especially rich in opportunities for functional combinations. The processes of life are chemical processes, so our `life is information' remains compatible with Kornberg's `life is chemistry', but goes deeper by highlighting the informational basis of the chemistry of life. Our perspective also emphasises the idea that the whole of life at all scales has a role in reproducing life. Considering life as information processing (computation) where the subject of computation is life, we are faced with a `program' running on itself, the function of which is to output itself. Such recursion is familiar and much exploited in computer science. It highlights the fact that for life, there is no distinction between the `machine' and the program - both are information; they are the same information, ordering and re-ordering matter and energy so as to persist. It would not be right to think of life as a biochemical structure on which a program is run, because life is the program and the biochemical structure is its embodiment. This is why we say that information is not just in DNA, but is in the whole biological system. The idea that `life is information processing' brings reductionists and synthesists closer together as it shows life to supervene on chemistry strictly according to information content, but to also possess strictly emergent properties (at several levels) arising from the \emph{functions} of the embodied information. Now that functional information content can be quantified at every level of life, we anticipate its use in further deepening our understanding of life and its place in the physical universe.

\subsection*{Acknowledgments}
This work was enhanced by very thoughtful and creative reviews by anonymous referees. It was supported by a Science Technology Research and Innovation for the Environment grant from the Environmental Protection Agency of the Republic of Ireland: 2007-PhD-SD-3. C.G. was partially supported by SNI membership 47907 of CONACyT, Mexico. 

\clearpage

\bibliographystyle{apalike}

\bibliography{BiodiversityValue33short}

\begin{thebibliography}{}

\bibitem[Adami et~al., 2000]{Adami:2000oq}
Adami, C., Ofria, C., and Collier, T.~C. (2000).
\newblock Evolution of biological complexity.
\newblock {\em Proc Natl Acad Sci U S A}, 97(9):4463--4468.

\bibitem[Annila and Annila, 2008]{Annila2008}
Annila, A. and Annila, E. ({2008}).
\newblock {Why did life emerge?}
\newblock {\em International Journal of Astrobiology}, {7}({3-4}):{293--300}.

\bibitem[Annila and Kuismanen, 2009]{Annila:2009kl}
Annila, A. and Kuismanen, E. (2009).
\newblock Natural hierarchy emerges from energy dispersal.
\newblock {\em Biosystems}, 95(3):227--233.

\bibitem[Balleza et~al., 2008]{Balleza:2008ix}
Balleza, E., Alvarez-Buylla, E.~R., Chaos, A., Kauffman, S., Shmulevich, I.,
  and Aldana, M. (2008).
\newblock Critical dynamics in genetic regulatory networks: Examples from four
  kingdoms.
\newblock {\em Plos One}, 3(6):e2456.

\bibitem[Bates, 2005]{Bates2005}
Bates, M. (2005).
\newblock Information and knowledge: an evolutionary framework for information
  science.
\newblock {\em Information Research}, 10(4):paper 239.

\bibitem[Bateson, 1972]{Bateson1972}
Bateson, G. (1972).
\newblock Form, substance, and difference.
\newblock In Bateson, G., editor, {\em Steps to an Ecology of Mind}, pages
  448--466. University of Chicago Press.

\bibitem[Bitbol and Luisi, 2004]{Bitbol:2004tv}
Bitbol, M. and Luisi, P. (2004).
\newblock Autopoiesis with or without cognition: defining life at its edge.
\newblock {\em J Royal Soc Interface}, 1(1):99--107.

\bibitem[Bowen and Jordan, 2002]{BowenJordan2002}
Bowen, N. and Jordan, I. (2002).
\newblock Transposable elements and the evolution of eukaryotic complexity.
\newblock {\em Curr Issues Mol Biol}, 4:65--76.

\bibitem[Bray, 1995]{BRAY:1995ph}
Bray, D. (1995).
\newblock Protein molecules as computational elements in living cells.
\newblock {\em Nature}, 376(6538):307--312.

\bibitem[Bray, 2009]{Bray2009}
Bray, D. (2009).
\newblock {\em Wetware: A computer in every living cell}.
\newblock Yale University Press, New Haven, CT. USA.

\bibitem[Butler et~al., 1998]{Paton1998}
Butler, M.~H., Paton, R.~C., and Leng, P.~H. (1998).
\newblock {\em Information processing in tissues and cells}, chapter
  Information processing in computational tissues, pages 177--184.
\newblock Plenum Press, New York.

\bibitem[Cairns et~al., 1988]{Cairns1988}
Cairns, J., Overbaugh, J., and Miller, S. (1988).
\newblock The origin of mutants.
\newblock {\em Nature}, 335:142--145.

\bibitem[Camazine et~al., 2001]{Camazinebook2001}
Camazine, S., Deneubourg, J.~L., Franks, N., Sneyd, J., Theraulaz, G., and
  Bonabeau, E. (2001).
\newblock {\em Self-Organization in Biological Systems}.
\newblock Princeton University Press, Princeton, NJ. USA.

\bibitem[Carroll, 2001]{Carroll:2001fe}
Carroll, S. (2001).
\newblock Chance and necessity: the evolution of morphological complexity and
  diversity.
\newblock {\em Nature}, 409(6823):1102--1109.

\bibitem[Chaitin, 1990]{Chaitin1990}
Chaitin, G. (1990).
\newblock {\em Information, Randomness and Incompleteness - Papers on
  Algorithmic Information Theory}, volume~8 of {\em Series in Computer
  Science}.
\newblock World Scientific, Singapore, 2nd edition.

\bibitem[Conant and Ashby, 1970]{ConantAshby1970}
Conant, R. and Ashby, W. (1970).
\newblock Every good regulator of a system must be a model of that system.
\newblock {\em International Journal of Systems Science}, 1(2):89--97.

\bibitem[Cowling et~al., 2004]{Cowling2004}
Cowling, R., Knight, A., Faith, D., Ferrier, S., Lombard, A., Driver, A.,
  Rouget, M., Maze, K., and Desmet, P. (2004).
\newblock Nature conservation requires more than a passion for species.
\newblock {\em Conserv Biol}, 18(6):1674--1676.

\bibitem[Cummins, 1975]{Cummins75}
Cummins, R. (1975).
\newblock Functional analysis.
\newblock {\em J. Philos.}, 72(20):741--765.

\bibitem[Curtis et~al., 2002]{Curtis:2002uk}
Curtis, T., Sloan, W., and Scannell, J. (2002).
\newblock Estimating prokaryotic diversity and its limits.
\newblock {\em Proc Natl Acad Sci U S A}, 99(16):10494--10499.

\bibitem[Davidson, 2001]{Davidson:2001aq}
Davidson, E.~H. (2001).
\newblock {\em Genomic regulatory systems: Development and evolution.}
\newblock Academic Press, San Diego, USA.

\bibitem[Davidson and Levin, 2005]{DavidsonLevin2005}
Davidson, E.~H. and Levin, M. (2005).
\newblock Gene regulatory networks.
\newblock {\em Proc Nat Acad Sci USA}, 102(14):4935.

\bibitem[Denning, 2007]{Denning2007}
Denning, P.~J. (2007).
\newblock Computing is a natural science.
\newblock {\em Communications of the ACM}, 50(7):13--18.

\bibitem[Dunne et~al., 2002]{Dunne:2002qq}
Dunne, J., Williams, R., and Martinez, N. (2002).
\newblock Food-web structure and network theory: The role of connectance and
  size.
\newblock {\em Proc Natl Acad Sci U S A}, 99(20):12917--12922.

\bibitem[Faeder, 2011]{Faeder:2011lr}
Faeder, J.~R. (2011).
\newblock Toward a comprehensive language for biological systems.
\newblock {\em BMC Biol}, 9:68.

\bibitem[Farnsworth et~al., 2012]{FarnsworthLyashevska2012}
Farnsworth, K., Lyashevska, O., and Fung, T. (2012).
\newblock Functional complexity: The source of value in biodiversity.
\newblock {\em Ecol Complex}, 11:46--52.

\bibitem[Favareau, 2009]{Favareau2009}
Favareau, D., editor (2009).
\newblock {\em Essential Readings in Biosemiotics: Anthology and Commentary}.
\newblock Springer, Berlin.

\bibitem[Floridi, 2003]{Floridi2003}
Floridi, L. (2003).
\newblock Information.
\newblock In Floridi, L., editor, {\em The Blackwell Guide to the Philosophy of
  Computing and Information}, pages 40--61. Blackwell Publishing Ltd.

\bibitem[Floridi, 2005]{Floridi:2005kx}
Floridi, L. (2005).
\newblock Is semantic information meaningful data?
\newblock {\em Philosophy and Phenomenological Research}, 70(2):351--370.

\bibitem[Galtin, 1972]{Galtin1972}
Galtin, L.~L. (1972).
\newblock {\em Information Theory and the Living System}.
\newblock Columbia University Press, New York.

\bibitem[Geard and Wiles, 2005]{Geard2005a}
Geard, N. and Wiles, J. (2005).
\newblock A gene network model for developing cell lineages.
\newblock {\em Artif Life}, 11:249--267.

\bibitem[Gell-Mann and Lloyd, 1996]{GellMann1996}
Gell-Mann, M. and Lloyd, S. (1996).
\newblock Information measures, effective complexity, and total information.
\newblock {\em Complexity}, 2(1):44--52.

\bibitem[Gell-Mann and Lloyd, 2003]{GellMann2003}
Gell-Mann, M. and Lloyd, S. (2003).
\newblock Effective complexity.
\newblock In Gell-Mann, M. and Tsallis, C., editors, {\em Nonextensive Entropy
  - Interdisciplinary Applications}. Oxford University Press.

\bibitem[Gershenson, 2010]{Gershenson:2010b}
Gershenson, C. (2010).
\newblock Computing networks: A general framework to contrast neural and swarm
  cognitions.
\newblock {\em Paladyn, Journal of Behavioral Robotics}, 1(2):147--153.

\bibitem[Gershenson and Fern\'andez, 2012]{GershensonFernandez:2012}
Gershenson, C. and Fern\'andez, N. (2012).
\newblock Complexity and information: Measuring emergence, self-organization,
  and homeostasis at multiple scales.
\newblock {\em Complexity}, Early View.

\bibitem[Goldman et~al., 1973]{Goldman1973}
Goldman, R., Pollack, R., and Hopkins, N. (1973).
\newblock Preservation of normal behavior by enucleated cells in culture.
\newblock {\em Proc Nat Acad Sci USA}, 70:750--754.

\bibitem[Gregory, 2001]{Gregory2001}
Gregory, T. (2001).
\newblock Coincidence, coevolution, or causation? {DNA} content, cell size, and
  the {C}-value enigma.
\newblock {\em Biological Reviews}, 76(1):65--101.

\bibitem[Griffiths, 1993]{Griffiths93}
Griffiths, P.~E. (1993).
\newblock Functional analysis and proper functions.
\newblock {\em British J. Philos. Sci.}, 44:409--422.

\bibitem[Hinegardner and Engelberg, 1983]{Hinegardner83a}
Hinegardner, R. and Engelberg, J. (1983).
\newblock Biological complexity.
\newblock {\em J Theor Biol}, 104:7--20.

\bibitem[Hopfield, 1994]{Hopfield1994}
Hopfield, J.~J. (1994).
\newblock Physics, computation, and why biology looks so different.
\newblock {\em J Theor Biol}, 171:53--60.

\bibitem[Jiang and Xu, 2010]{Jiang:2010lr}
Jiang, Y. and Xu, C. (2010).
\newblock The calculation of information and organismal complexity.
\newblock {\em Biol Direct}, 5:59.

\bibitem[Kaila and Annila, 2008]{Kaila:2008pb}
Kaila, V. R.~I. and Annila, A. (2008).
\newblock Natural selection for least action.
\newblock {\em Proceedings of the Royal Society A-Mathematical Physical and
  Engineering Sciences}, 464(2099):3055--3070.

\bibitem[Karnani and Annila, 2009]{Karnani:2009rq}
Karnani, M. and Annila, A. (2009).
\newblock Gaia again.
\newblock {\em Biosystems}, 95(1):82--87.

\bibitem[Kauffman, 1993]{Kauffman93}
Kauffman, S.~A. (1993).
\newblock {\em Origins of Order: Self-Organization and Selection in Evolution}.
\newblock Oxford University Press, Oxford, UK.

\bibitem[Kohl et~al., 2010]{Kohl:2010fm}
Kohl, P., Crampin, E.~J., Quinn, T.~A., and Noble, D. (2010).
\newblock Systems biology: An approach.
\newblock {\em Clin Pharmacol Ther}, 88(1):25--33.

\bibitem[Kornberg, 1991]{KORNBERG:1991rz}
Kornberg, A. (1991).
\newblock Understanding life as chemistry.
\newblock {\em Clin Chem}, 37(11):1895--1899.

\bibitem[Kravchenko-Balasha et~al., 2012]{Kravchenko-Balasha:2012tb}
Kravchenko-Balasha, N., Levitzki, A., Goldstein, A., Rotter, V., Gross, A.,
  Remacle, F., and Levine, R.~D. (2012).
\newblock On a fundamental structure of gene networks in living cells.
\newblock {\em Proc Natl Acad Sci U S A}, 109(12):4702--7.

\bibitem[Lee, 2004]{Lee2004}
Lee, K. (2004).
\newblock There is biodiversity and biodiversity.
\newblock In Oksanen, M. and Pietarinen, J., editors, {\em Philosophy and
  Biodiversity}, pages 152--171. Cambridge University Press, Cambridge, UK.

\bibitem[Lehn, 1990]{Lehn:1990}
Lehn, J.-M. (1990).
\newblock Perspectives in supramolecular chemistry---from molecular recognition
  towards molecular information processing and self-organization.
\newblock {\em Angewandte Chemie International Edition in English},
  29(11):1304--1319.

\bibitem[Lepot et~al., 2008]{Lepot2008}
Lepot, K., Benzerara, K., Brown, G., and P, P. (2008).
\newblock Microbially influenced formation of 2,724-million-year-old
  stromatolites.
\newblock {\em Nat. Geosci.}, 1:118--121.

\bibitem[Li and Vit{\'a}nyi, 2008]{Li2008}
Li, M. and Vit{\'a}nyi, P. M.~B. (2008).
\newblock {\em An introduction to Kolmogorov complexity and its applications}.
\newblock Springer, 3rd edition.

\bibitem[Lorenz et~al., 2011]{Lorenz2011129}
Lorenz, D.~M., Jeng, A., and Deem, M.~W. (2011).
\newblock The emergence of modularity in biological systems.
\newblock {\em Physics of Life Reviews}, 8(2):129 -- 160.

\bibitem[Lovelock and Margulis, 1974]{Lovelock1974}
Lovelock, J.~E. and Margulis, L. (1974).
\newblock Atmospheric homeostasis by and for the biosphere: {The Gaia}
  hypothesis.
\newblock {\em Tellus}, 26(1):2--10.

\bibitem[Lyashevska and Farnsworth, 2012]{Lyashevska2012}
Lyashevska, O. and Farnsworth, K.~D. (2012).
\newblock How many dimensions of biodiversity do we need?
\newblock {\em Ecological Indicators}, 18:485--492.

\bibitem[MacKay, 1969]{MacKay1969}
MacKay, D.~M. (1969).
\newblock {\em Information, Mechanism and Meaning}.
\newblock MIT Press, Cambridge, MA, USA.

\bibitem[Magurran, 2004]{magurran2004}
Magurran, A. (2004).
\newblock {\em Measuring Biological Diversity}.
\newblock Blackwell Publishing.

\bibitem[Margulis, 1970]{Margulis1970}
Margulis, L. (1970).
\newblock {\em Origin of Eukaryotic Cells}.
\newblock Yale University Press, New Haven, CT. USA.

\bibitem[Maturana and Varela, 1980]{Maturana1980}
Maturana, H. and Varela, F.~J. (1980).
\newblock {\em Autopoiesis and Cognition: the Realization of the Living.}
\newblock D. Reidel Publishing Company, Dordrecht, NL.
\newblock Translation of original: De Maquinas y seres vivos. Universitaria
  Santiago.

\bibitem[Maus et~al., 2011]{Maus:2011fk}
Maus, C., Rybacki, S., and Uhrmacher, A.~M. (2011).
\newblock Rule-based multi-level modeling of cell biological systems.
\newblock {\em BMC Syst Biol}, 5:166.

\bibitem[McAllister, 2003]{McAllister2003}
McAllister, J. (2003).
\newblock Effective complexity as a measure of information content.
\newblock {\em Philos Sci}, 70(2):302--307.

\bibitem[McGill, 2011]{McGill2011}
McGill, B.~J. (2011).
\newblock Linking biodiversity patterns by autocorrelated random sampling.
\newblock {\em Am J Bot}, 98(3):481--502.

\bibitem[Menconi, 2005]{Menconi2005}
Menconi, G. (2005).
\newblock Sublinear growth of information in dna sequences.
\newblock {\em Bull. Math. Biol.}, 67(4):737--759.

\bibitem[Montoya et~al., 2006]{Montoya2006}
Montoya, J., Pimm, S.~L., and Sol{\'e}, R.~V. (2006).
\newblock Ecological networks and their fragility.
\newblock {\em Nature}, 442(7100):259--264.

\bibitem[Mora et~al., 2011]{Mora:2011eu}
Mora, C., Tittensor, D.~P., Adl, S., Simpson, A. G.~B., and Worm, B. (2011).
\newblock How many species are there on earth and in the ocean?
\newblock {\em PLoS Biol}, 9(8):e1001127.

\bibitem[Morowitz, 1992]{Morowitz1992}
Morowitz, H.~J. (1992).
\newblock {\em Beginnings of Cellular Life}.
\newblock Yale University Press, New Haven, CT. USA.

\bibitem[Neander, 1991]{Neander91}
Neander, K. (1991).
\newblock Functions as selected effects: A conceptual analysts defense.
\newblock {\em Philos. Sci.}, 58(2):168--184.

\bibitem[Neander, 2011]{Neander2011}
Neander, K. (2011).
\newblock {\em Routledge Encyclopedia of Philosophy (Online)}.
\newblock Routledge.

\bibitem[Nekola and White, 1999]{Nekola1999}
Nekola, J. and White, P. (1999).
\newblock The distance decay of similarity in biogeography and ecology.
\newblock {\em J Biogeogr}, 26(4):867--878.

\bibitem[Norton and Ulanowicz, 1992]{NORTON:1992tx}
Norton, B. and Ulanowicz, R. (1992).
\newblock Scale and biodiversity policy - a hierarchical approach.
\newblock {\em Ambio}, 21(3):244--249.

\bibitem[Orchard et~al., 2005]{Orchard:2005qd}
Orchard, S., Hermjakob, H., and Apweiler, R. (2005).
\newblock Annotating the human proteome.
\newblock {\em Mol Cell Proteomics}, 4(4):435--40.

\bibitem[Orgel and Crick, 1980]{Orgel1980}
Orgel, L. and Crick, F. (1980).
\newblock Selfish {DNA}: {The} ultimate parasite.
\newblock {\em Nature}, 284:604--607.

\bibitem[Prigogine, 1977]{Prigogine77}
Prigogine, I. (1977).
\newblock {\em Self-Organization in Non-Equilibrium Systems}.
\newblock Wiley, New York.

\bibitem[Prigogine and Stengers, 1984]{Prigogine84}
Prigogine, I. and Stengers, I. (1984).
\newblock {\em Order out of Chaos: Man's new dialogue with nature.}
\newblock Flamingo. Collins Publishing Group., London.

\bibitem[Robertson and Joyce, 2010]{RobertsonJoyce2010}
Robertson, M. and Joyce, G. (2010).
\newblock The origins of the rna world.
\newblock {\em Cold Spring Harbour Perspectives In Biology}.

\bibitem[Rodbell, 1995]{RodbellM1995}
Rodbell, M. (1995).
\newblock Signal transduction: evolution of an idea.
\newblock {\em Biosci Rep.}, 15:117--133.

\bibitem[Salthe, 1985]{salthe1985}
Salthe, S. (1985).
\newblock {\em Evolving Hierarchical Systems: Their Structure and
  Representation}.
\newblock Columbia University Press.

\bibitem[Schneider, 2000]{Schneider.ev2000}
Schneider, T.~D. (2000).
\newblock Evolution of biological information.
\newblock {\em Nucleic Acids Res}, 28:2794--2799.

\bibitem[Schr\"{o}dinger, 1944]{Schrodinger1944}
Schr\"{o}dinger, E. (1944).
\newblock {What is Life? The physical aspects of the living cell}.
\newblock http://home.att.net/~p.caimi/schrodinger.html.

\bibitem[Scott and Carr, 1992]{Scott1992}
Scott, J. and Carr, W. (1992).
\newblock Subcellular localization of the type {II cAMP-dependent} protein
  kinase.
\newblock {\em Physiology}, 7:143--148.

\bibitem[Shannon, 1948]{Shannon1948}
Shannon, C. (1948).
\newblock A mathematical theory of communication.
\newblock {\em Bell System Technical Journal}, 27(3,4):379--423,623--656.

\bibitem[Smith, 2008]{Smith:2008hy}
Smith, E. (2008).
\newblock Thermodynamics of natural selection i: Energy flow and the limits on
  organization.
\newblock {\em J Theor Biol}, 252(2):185--197.

\bibitem[Smith and Morowitz, 2004]{SmithMorowitz2004}
Smith, E. and Morowitz, H.~J. (2004).
\newblock Universality in intermediary metabolism.
\newblock {\em Proc Nat Acad Sci USA}, 101(36):13168--13173.

\bibitem[Szostak, 2003]{Szostak:2003kl}
Szostak, J.~W. (2003).
\newblock Functional information: Molecular messages.
\newblock {\em Nature}, 423(6941):689--689.

\bibitem[Tuomisto, 2010]{Tuomisto12010}
Tuomisto, H. (2010).
\newblock A diversity of beta diversities: straightening up a concept gone
  awry. part 1. defining beta diversity as a function of alpha and gamma
  diversity.
\newblock {\em Ecography}, 33(1):2--22.

\bibitem[Turing, 1936]{Turing1936}
Turing, A. (1936).
\newblock On computable numbers, with an application to the entscheidungs
  problem.
\newblock {\em Proceedings of the London Mathematical Society}, 42:230--265.

\bibitem[Turing, 1952]{Turing1952}
Turing, A. (1952).
\newblock The chemical basis for morphogenesis.
\newblock {\em Philos. Trans. R. Soc. Lond. Ser. B-Biol. Sci.}, 237:37--72.

\bibitem[Ulanowicz, 1980]{Ulanowicz1980}
Ulanowicz, R. (1980).
\newblock An hypothesis on the development of natural communitiesl.
\newblock {\em J Theor Biol}, 85:223--245.

\bibitem[Ulanowicz and Baird, 1999]{Ulanowicz1999}
Ulanowicz, R. and Baird, D. (1999).
\newblock Nutrient controls on ecosystem dynamics: The chesapeake mesohaline
  community.
\newblock {\em J Mar Syst}, 19:159--172.

\bibitem[Valentine, 1994]{VALENTINE:1994hi}
Valentine, J. (1994).
\newblock Late precambrian bilaterians: grades and clades.
\newblock {\em Proc Natl Acad Sci U S A}, 91(15):6751--6757.

\bibitem[Valentine, 2003a]{Valentine2003}
Valentine, J. (2003a).
\newblock Architectures of biological complexity.
\newblock {\em Integrative and comparative biology}, 43(1):99--103.

\bibitem[Valentine, 2003b]{Valentine2003bk}
Valentine, J. (2003b).
\newblock Cell types, cell type numbers, and body plan complexity.
\newblock In Hall, B. and Olson, W., editors, {\em Keywords and Concepts in
  Evolutionary Developmental Biology}, pages 35--43. Harvard University Press,
  Cambridge, MA, USA.

\bibitem[Valentine et~al., 1994]{VALENTINE:1994kn}
Valentine, J., Collins, A., and Meyer, C. (1994).
\newblock Morphological complexity increase in metazoans.
\newblock {\em Paleobiology}, 20(2):131--142.

\bibitem[Veitia and Bottani, 2009]{Veitia:2009zz}
Veitia, R.~A. and Bottani, S. (2009).
\newblock Whole genome duplications and a `function' for junk {DNA}? {Facts}
  and hypotheses.
\newblock {\em Plos ONE}, 4(12):e8201.

\bibitem[von Foerster, 1960]{Foerster60}
von Foerster, H. (1960).
\newblock On self-organizing systems and their environments.
\newblock In Yovits, M. and Cameron, S., editors, {\em Self-organizing
  systems}. Pergamon Press, Oxford, UK.

\bibitem[Wessler, 2006]{Wessler:2006ls}
Wessler, S.~R. (2006).
\newblock Transposable elements and the evolution of eukaryotic genomes.
\newblock {\em Proc Natl Acad Sci U S A}, 103(47):17600--17601.

\bibitem[Wicken, 1979]{Wicken1979}
Wicken, J.~S. (1979).
\newblock The generation of complexity in evolution: A thermodynamic and
  information-theoretical discussion.
\newblock {\em J Theor Biol}, 77:349--365.

\bibitem[Wiener, 1948]{Wiener1948}
Wiener, N. (1948).
\newblock {\em Cybernetics; or, Control and Communication in the Animal and the
  Machine.}
\newblock Wiley and Sons, New York.

\bibitem[Yockey et~al., 1958]{Yockey1958}
Yockey, H., Platzman, R., and Quastler, H., editors (1958).
\newblock {\em Symposium on Information Theory in Biology (1956 : Gatlinburg,
  Tenn.)}.
\newblock Pergamon Press, New York.

\bibitem[Zepik et~al., 2001]{Zepik:2001ta}
Zepik, H., Blochliger, E., and Luisi, P. (2001).
\newblock A chemical model of homeostasis.
\newblock {\em Angewandte Chemie-International Edition}, 40(1):199--202.

\bibitem[Zhou et~al., 2010]{Zhou:2010fj}
Zhou, J., Deng, Y., Luo, F., He, Z., Tu, Q., and Zhi, X. (2010).
\newblock Functional molecular ecological networks.
\newblock {\em MBio}, 1(4):e00169--00110.

\end{thebibliography}

\clearpage

\begin{table}[h!]
\centering

\label{tab:hierarchy}
\begin{tabular}{l|l}

\hline
\textbf{Organization Level} & \textbf{Interactions} \\ 
\hline
life as a whole & global bio-geochemical networks   \\ \hline
ecological communities & interspecific material and energy flows \\ \hline
populations - species & gene-flow, dispersal, evolution  \\ \hline
multi-cellular organisms & organism societies + interspecific, e.g. parasitism \\ \hline
tissues, organs and organ systems & cellular communication and organ function \\ \hline
cells & specialisation and ontogeny: e.g. immune system \\ \hline
sub-cellular structures & catabolic autopoietic processes \\ \hline
molecular networks & metabolic and information processing \\ \hline
DNA sequences: codons to genes & coding and expression control \\ \hline
molecular surfaces & lock and key - enzymes \\ \hline
\end{tabular}

\caption{A ten-level hierarchy of biocomplexity. Left column names the level of organization and right column gives examples of the complex interactions and processes that take place at that level, contributing to biocomplexity. Complexity is also added by interactions among levels, both upwards and downwards, producing feedback circuits. Interactions at every level and among levels constitute information processing.  (adapted from \cite{FarnsworthLyashevska2012})} 
\end{table}

\end{document}